\documentclass[10pt,twocolumn,amsmath,amssymb,floatfix,showpacs]{revtex4}
\usepackage{graphicx,epsfig,latexsym,amssymb}
\usepackage{multirow,amsmath,array,booktabs}
\usepackage{dcolumn}
\usepackage{color}
\usepackage{textcomp}
\usepackage{longtable}
\usepackage[section]{placeins}
\usepackage{bm}

\usepackage{color}
\usepackage{CJK}
\usepackage{graphicx}% Include figure files
\usepackage{dcolumn}% Align table columns on decimal point
\usepackage{bm}% bold math
\usepackage{epstopdf}
\usepackage{multirow}
\usepackage{booktabs}
\usepackage{subfigure}
\usepackage[dvipdfmx,bookmarks=true,colorlinks,%
citecolor=blue,linkcolor=blue,anchorcolor=blue,filecolor=blue,urlcolor=blue,%
]{hyperref}
\usepackage{ulem}
\usepackage{titlesec}
\titlespacing{\section}{0pt}{10pt}{10pt}[0pt]

\begin{document}

	\title{Study on deformed halo nucleus $^{31}$Ne with Glauber model based \\
	on  microscopic self-consistent structures}
	\hyphenpenalty=20000
	\tolerance=2000
\author{Shi-Yi Zhong$^{1}$}
\author{Shi-Sheng Zhang$^{1}$}\email{zss76@buaa.edu.cn}
\author{Xiang-Xiang Sun$^{2}$}\email{sunxiangxiang@ucas.ac.cn}
\author{Michael S. Smith$^{3}$}\email{smithms@ornl.gov}
\affiliation{$^{1}$School of Physics, Beihang University, Beijing 100191, China}
\affiliation{$^{2}$School of Nuclear Science and Technology,
			University of Chinese Academy of Sciences,
			Beijing 100049, China}
\affiliation{$^{3}$Physics Division, Oak Ridge National Laboratory, Oak Ridge, Tennessee, 37831-6354, USA}

		\date{\today}
		\begin{abstract}			
% note that pairing and resonances are mentioned here in the abstract but not at all in the paper
We study the exotic deformed nucleus $^{31}$Ne using an approach that combines self-consistent structure and reaction theory. We utilize the fully-relativistic, microscopic deformed Hartree-Bogoliubov theory in continuum (DRHBc) to demonstrate that deformation and pairing correlations give rise to a halo structure with large-amplitude $p$-wave configuration in $^{31}$Ne. We then use the valence nucleon wave functions and angle-averaged density distributions of $^{30}$Ne from this model as input for a Glauber reaction model to study the observables of neutron-rich Neon isotopes and search for halo signatures. Our predictions of the reaction cross sections of these exotic Neon isotopes on a Carbon target can better reproduce the experimental data than those from relativistic mean field model for a spherical shape with resonances and pairing correlations contributions, as well as those from a Skyrme-Hartree-Fock model. The one-neutron removal cross section at 240 MeV/nucleon, the inclusive longitudinal momentum distribution of the $^{30}$Ne, and the valence neutron residues from the $^{31}$Ne breakup reaction are largely improved over previous theoretical predictions and agree well with data. These reaction data indicate a dilute density distribution in coordinate space and are a canonical signature of a halo structure.
		\end{abstract}
		\maketitle

%%introduction%%%%%%%%%%%%%%%%%%%%%%%%%%%%%%
\section{INTRODUCTION}
Of the 3000 nuclei that can be reached by experimentalists and the 9000 predicted by theory to be bound~\cite{Erler2012,XIA2018_ADNDT121}, less than 30 are confirmed as ``halo'' nuclei with one or more weakly bound valence nucleon(s) exhibiting a significantly extended spatial distribution. Interest in discovering more of these rare nuclei, and understanding the mechanism of their formation, has been high since the 1985 discovery of the first halo nucleus, $^{11}$Li~\cite{Tanihata1985_PRL55-2676}. For heavier nuclei, structure effects including configuration mixing, coupling to the continuum, pairing correlations, and deformation may all influence the formation of nuclear halos~\cite{Meng2015_JPG093101}.  

Studies of halo nuclei closely couple structure and reaction aspects, and an important goal is a self-consistent approach wherein a structure model predicts halo \textit{characteristics}, which are then input into a reaction model to predict \textit{observable signatures} of a halo that can be compared to measurements. The structure characteristics include an extended matter radius, an extended neutron density distribution, and occupation probabilities of valence orbitals, while reaction signatures include an enhanced total reaction cross section, enhanced Coulomb breakup cross section, enhanced nucleon removal cross section, and a peaked momentum distribution (dilute spatial distribution) following nuclear breakup~\cite{Al-Khalili2004_LNP651-77,Tanihata2013_PPNP215}.

A measurement in 2009 of the large Coulomb breakup cross section of $^{31}$Ne bombarding Pb and C targets, which showed a large soft $E$1 excitation for the ground state, was interpreted as a $p$-wave halo~\cite{nakamura2009halo} -- making $^{31}$Ne the heaviest halo nucleus found at that time. Subsequent measurements of the interaction cross section found an enhancement of 12$\%$ above those of neighboring Neon isotopes, which suggested an $s$- or $p$- wave halo in $^{31}$Ne~\cite{Takechi2012_PLB707-357}. 

A number of theoretical studies of halo phenomena in Neon isotopes followed~\cite{urata2012reaction,  Minomo2012_PRL108-052503, horiuchi2012glauber}. In 2014, we utilized the analytical continuation of the coupling constant (ACCC) method, based on a relativistic mean field (RMF) structure theory with the Bardeen-Cooper-Schrieffer (BCS) pairing approximation -- the RAB model~\cite{Zhang2014_PLB730-30} -- to calculate neutron- and matter-radii, one-neutron separation energies, $p$- and $f$-wave energies and occupation probabilities,
and neutron densities for single-particle resonant orbitals in $^{27-31}$Ne. These results were analyzed for characteristics of neutron halo formation in $^{31}$Ne due to a competition of the occupation for unbound resonant orbitals and the pairing correlations. Based on a radius increase from $^{30}$Ne to $^{31}$Ne that is much larger than the increase from $^{29}$Ne to $^{30}$Ne, and a simultaneous decrease in the neutron separation energy, a $p$-wave 1$n$ halo structure was predicted for $^{31}$Ne. Later that year, the spin-parity 3/2$^{-}$ of its valence neutrons were confirmed from experiment~\cite{Nakamura2014_PRL112-142501}, consistent with our predictions, and $^{31}$Ne was confirmed to have all the characteristics of a $p$-wave halo structure.

A number of previous attempts to quantitatively explain the trends in $^{31}$Ne reaction observables with theoretical reaction models, without any additional effect, have had only limited success. For example, structure input from a non-relativistic mean field Skyrme-Hartree-Fock (SHF) model were used~\cite{horiuchi2012glauber} as input for the Glauber reaction model, but could not reproduce the measured large increase in the interaction cross section at $^{31}$Ne (compared to $^{30}$Ne). The lack of contributions from pairing correlations and the continuum in the SHF model may have caused this underestimate. In Ref.~\cite{sumi2012deformation}, an Antisymmetrized Molecular Dynamics (AMD) approach was used with a double-folding model to predict $^{31}$Ne + $^{12}$C $\to$ $^{30}$Ne + X reaction cross sections, but largely underestimated the reaction cross section increase. The further addition of a Resonating Group Method (RGM), however, partially improved issues of the density ``tail'', enabling reaction cross section predictions to be approximately 1$\sigma$ below the measured data. 

Recently, we used structure predictions from the RAB model, which includes the assumption of a spherical nucleus shape, as input for a Glauber reaction model and examined the predictions for halo signatures~\cite{Zhang2021_arXiv2108.05609}. Our predictions of total reaction and one-neutron removal cross sections of $^{31}$Ne on a Carbon target were significantly enhanced compared with those of neighboring Neon isotopes, agreeing with measurements at 240 MeV/nucleon and consistent with a single neutron halo. Furthermore, our RAB calculations of the inclusive longitudinal momentum distribution of the $^{30}$Ne and valence neutron residues from the $^{31}$Ne breakup reaction indicate a dilute density distribution in coordinate space.

The spherical nature of the RAB model in that study, however, could not account for any effects arising from the deformation of $^{31}$Ne. Experiments show $^{31}$Ne has relatively large deformation $\beta_{2}$~\cite{Nakamura2014_PRL112-142501}, thought to be between 0.2 and 0.42 in Refs.~\cite{urata2012reaction, sumi2012deformation, Minomo2012_PRL108-052503, horiuchi2012glauber, geng2004proton}. In this study, we use a deformed model to take into account the contribution of the deformation $\beta_{2}$ in our predictions of the structure properties of $^{31}$Ne, and then use these predictions in a Glauber reaction model to calculate halo observables and compare to measurements. Our structure model is the deformed relativistic Hartree-Bogoliubov theory in continuum (DRHBc)~\cite{Zhou2010_PRC82-011301R,Li2012_PRC85-024312}, a microcosmic self-consistent theory with discretized continuum contribution, deformation and pairing correlations by the Bogoliubov transform, which is based on spherical Woods-Saxon (WS) basis expansion~\cite{Zhou2003_PRC68-034323}. This is systematic structure study of the Neon isotopic chain with a deformed relativistic model coupled to a reaction model in a search for nuclear halo observables. 

This paper is made up of three sections. In Section~\ref{section 2}, the DRHBc theory is used to predict the structure properties of $^{26-31}$Ne. The single neutron orbitals, configurations, and density distributions of deformed $^{31}$Ne are analyzed for halo characteristics. The core density distribution of this nucleus is then averaged over angle for, along with the valence nucleon wave function, input into a Glauber reaction in Section~\ref{section 3}. The inclusive longitudinal momentum distribution after breakup, the one-neutron removal cross section, and the reaction cross section as a function of mass of Neon isotopes bombarding on a Carbon target are predicted with this model and compared with experimental data. Comparisons are also made with predictions using the spherical RAB model~\cite{Zhang2021_arXiv2108.05609} to determine the impact of deformation on reaction halo observables. A summary of the study is given in Section~\ref{section 4}.

\section{STRUCTURE STUDY OF DEFORMED EXOTIC NEON ISOTOPES}
\label{section 2}
In Ref.~\cite{Zhang2014_PLB730-30},
the halo structure of $^{31}$Ne in spherical limit has been studied by using the RAB model
with the contributions from resonant orbitals and pairing correlations with the effective interactions NL1~\cite{Reinhard1986_ZPA323-13} and NL3~\cite{Lalazissis97}.
To study the impacts of \textit{deformation} on the halo structure of $^{31}$Ne,
we adopt the DRHBc theory to explore the properties of the ground-states in $^{26-31}$Ne. We use the NL1 effective interaction in the particle-hole channel, and a density-dependent zero range pairing force in the particle-particle channel.
For odd mass nuclei, blocking effects are considered~\cite{Li2012_CPL29-042101}.
The DRHBc theory has been used to study deformed halos in
$^{17,19}$B \cite{Yang2021_PRL126-082501,Sun2021_PRC103-054315},
carbon isotopes \cite{Sun2018_PLB785-530,Sun2020_NPA1003-122011},
$^{36,38}$Ne \cite{Zhou2010_PRC82-011301R},
and $^{42,44}$Mg \cite{Zhou2010_PRC82-011301R,Li2012_PRC85-024312,Zhang2019_PRC100-034312,Sun2021_SciB2095}.
Recently, a nuclear mass table that includes deformation effects and the continuum is in
progress using this theory
\cite{Pan2019_IJMPE28-1950082,Zhang2020_PRC102-024314,In2021_IJMPE30-2150009,Pan2021_PRC024331}.
The detailed formulae of this theory can be found in Refs.~\cite{Li2012_PRC85-024312,Sun2021_SciB2095, Zhang2020_PRC102-024314} and references therein.

In this work, pairing strength is fixed to be 380 MeV fm$^{-3}$ connected with a smooth
energy cut-off in the quasi-particle space~\cite{Li2012_PRC85-024312}.
The box size used to generate the Dirac WS basis is 20 fm with a step size of 0.05 fm, and we include up to fourth-order Legendre polynomials. We have checked that suitable convergence can be achieved for all our calculations with these parameters.

In this scheme, the calculated rms matter radii are 3.29 fm, 3.34 fm, and 3.46 fm for $^{29,30,31}$Ne with the quadrupole
deformation parameter $\beta_2$ of 0.16, $-0.07$, and 0.26, respectively.
A sudden radius increase appears from $^{30}$Ne to $^{31}$Ne, which is 2.4 times larger than the increase from $^{29}$Ne to $^{30}$Ne.
Additionally, we predict a one-neutron separation energy $S_n$ for $^{31}$Ne of 1.03 MeV, a factor of 2.6 lower than that of $^{29}$Ne; this is in agreement with the (weak) constraint on that ratio from experiment of 5.7 $\pm$ 4.4  
~\cite{Wang2021_ChinPC45-030003}.
%%%%%%%%
%NOTE: The important point is that you see a significant decrease in the separation energy when going from 29Ne to 31Ne, both in the model and in the measured values. Since our predicted values are significantly higher than the measured values, I just thought we could quote the ratios and not the actual values.}
%%%%%%%%%%%%%%%%%%%%%

\begin{figure}[htb]
\begin{center}
\includegraphics[width=0.5\textwidth]{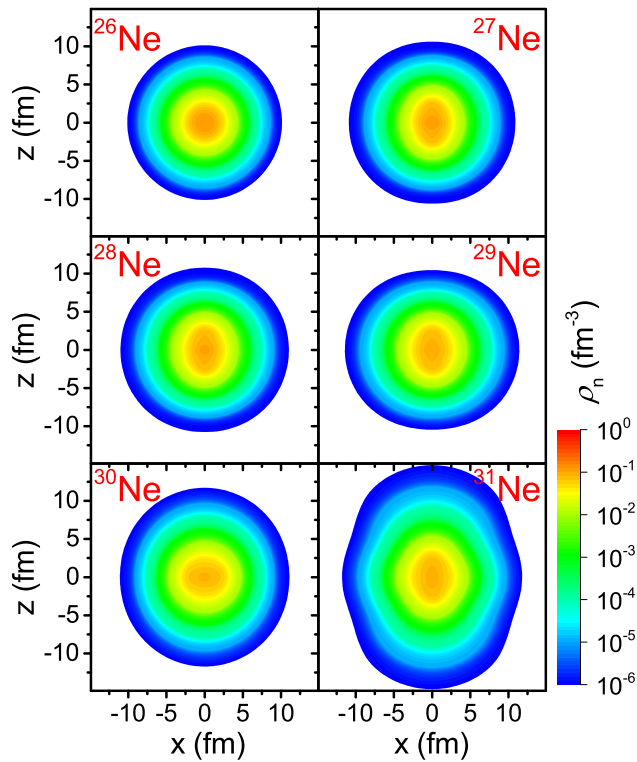}
\end{center}
\caption{
Two-dimensional neutron density distributions for $^{26-31}$Ne in $xz$ plane
from DRHBc calculations with the NL1 effective interaction.
The $z$-axis is the symmetry axis.}
\label{fig1}
\end{figure}

In Fig.~\ref{fig1}, we plot two-dimensional neutron density distributions in $^{26-31}$Ne.
It can be seen that $^{27,28,29,31}$Ne display a prolate shape
while $^{26,30}$Ne have a nearly spherical shape in our calculations.
For the nucleus $^{31}$Ne, the neutron density distribution is much wider than the other Neon isotopes, especially along the $z$-axis. This results in a 2.4 times larger increase in the neutron density distribution radius from $^{30}$Ne to $^{31}$Ne compared to the increase from $^{29}$Ne to $^{30}$Ne. The large spatial extension of neutrons, the small one-neutron separation energy,
and the enhanced rms matter radius all indicate that $^{31}$Ne is a halo nucleus.

\begin{figure}[htb]
	\begin{center}
		\includegraphics[width=0.5\textwidth]{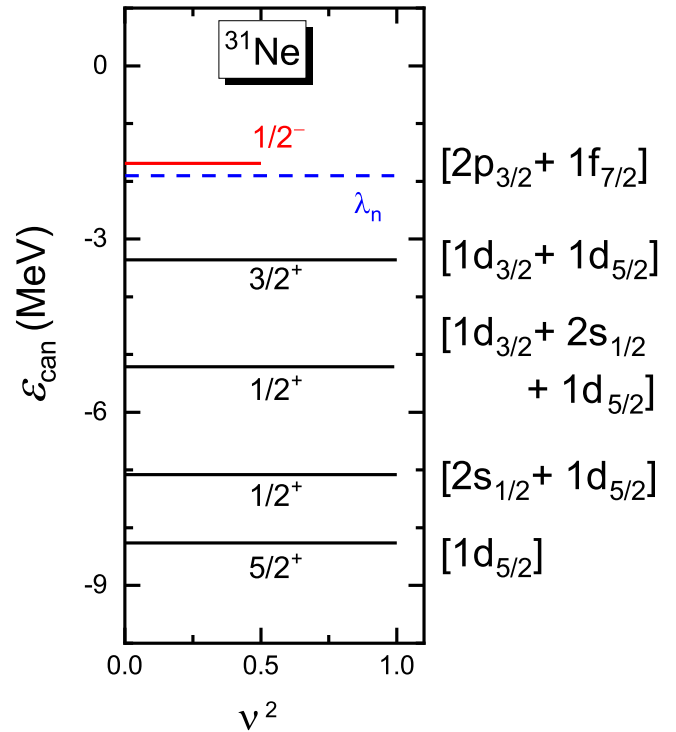}
	\end{center}
	\caption{
		Single neutron levels with the quantum
		numbers $m^\pi$ near the Fermi surface $\lambda_n$ (dashed line) in the
		canonical basis for $^{31}$Ne as a function of the occupation probability
		$v^2$. The main spherical components for orbitals that are close
		to the threshold are also given. The levels with positive (negative) parity
		are labeled by black (red) lines. The length of each solid line is equal
		to the value of $v^2$.
	}
	\label{fig2}
\end{figure}

\begin{figure}[htb]
\begin{center}
\includegraphics[width=0.5\textwidth]{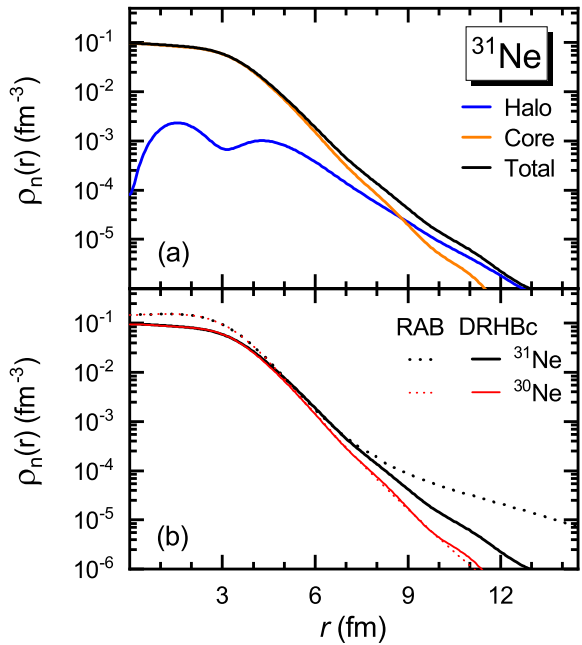}
\end{center}
\caption{
(a) Angle-averaged neutron densities in $^{31}$Ne (black solid line) for the valence (halo) contribution (blue solid line) and the core contribution (dark-yellow solid line) from the DRHBc theory with NL1 effective interaction. (b) The total angle-averaged neutron densities for $^{30}$Ne (red lines) and $^{31}$Ne (black lines) from the DRHBc theory (solid lines) and the spherical RAB model (dotted lines), respectively.
}
	
\label{fig3}
\end{figure}

To analyze the mechanism that gives rise to a halo, we display the single-particle levels (SPLs) of $^{31}$Ne in the canonical basis and the main spherical components of each orbital in Fig.~\ref{fig2}. The spin-parity for the valence neutron orbital is $1/2^-$ with the occupation probability of 0.5. The neutron Fermi energy $\lambda_n$ is $-1.8$ MeV, and the sizable energy gap between the $1/2^-$ and $3/2^+$ orbitals divides the neutrons of $^{31}$Ne into two parts: the $1/2^-$ valence orbital is responsible for the formation of the {\it{halo}}, while the remaining orbitals contribute to the {\it{core}}.
The configuration of the valence orbital is mainly from the mixture of 
$2p_{3/2}$, $3p_{3/2}$, $2p_{1/2}$, and $1f_{7/2}$ configurations, 
with contributions of $41.4\%$, $9.8\%$, $7.8\%$, and $35.3\%$, respectively.
To clearly see the contributions of these orbitals to the density distribution of $^{31}$Ne, in Fig.~\ref{fig3}\href{fig3}{(a)} we separately plot the valence contribution and core contribution, which sum up to the total angle-averaged density distribution of $^{31}$Ne; the angle averaging is necessary to serve as input for the spherical Glauber reaction model in Section~\ref{section 3}. From this plot, it is clear that the $p$-wave components in the weakly bound $1/2^-$ level contribute a density distribution that extends to a large radius, and therefore are responsible for the halo nature of $^{31}$Ne.

We note that the $1/2^-$ spin-parity of $^{31}$Ne in this work differs from
the $3/2^-$ value given in Refs.~\cite{Takechi2012_PLB707-357,Minomo2012_PRL108-052503,Nakamura2014_PRL112-142501}, which might be the consequence of our predictions of slightly smaller deformation. This does not, however, impact the conclusion that the halo in $^{31}$Ne is caused by a $p$-wave. In the study of halos in deformed nuclei, DRHBc theory is known to predict shape decoupling effects in certain nuclei, wherein the core and the halo have different shapes~\cite{Zhou2010_PRC82-011301R}. As for $^{31}$Ne, we find that both the core and the halo are prolate, therefore $^{31}$Ne is a deformed halo nuclei that does not display shape decoupling.

We also compare the angle-averaged density distributions of $^{30,31}$Ne from the DRHBc theory with those from the spherical RAB model~\cite{Zhang2014_PLB730-30} in Fig.~\ref{fig3}\href{fig3}{(b)}. While the  $^{31}$Ne density distribution from the DRHBc theory is somewhat denser than that from the RAB model, it is still dilute compared with the core nucleus $^{30}$Ne. The different predictions between the two methods for $^{31}$Ne arises from the different ordering of the orbitals in the spherical and deformed treatments. While the valence neutron occupies a resonant orbital above the threshold of neutron emission (\textit{i.e.}, with positive energy) in the RAB model, it occupies a weakly bound orbital \textit{below} the threshold of neutron emission in the DRHBc theory. 

In summary, our structure calculations of the deformed exotic nucleus $^{31}$Ne with the fully-relativistic, microscopic DRHBc structure theory predicts that a halo structure is formed from a low-angular momentum $p$-$wave$ configuration of the valence neutron. The halo characteristics are a significantly more dilute density distribution, a significantly smaller separation energy, and a significant increase in radius compared to neighboring Ne nuclei. The DRHBc theory predicts the valence neutron is in a weakly bound orbital with low-angular momentum configuration, in contrast to our previous spherical model calculations.
We also made calculations with the NL3 and PK1 effective interactions~\cite{Lalazissis97, long2004new} and found 
    a similar angle-averaged density distribution
    as that for the NL1 effective interaction, but with a slightly denser tail. 
    The deformations $\beta_2$ are 0.25, 
    slightly smaller than the previous prediction.

\section{REACTION STUDY OF EXOTIC NEON ISOTOPES}
\label{section 3}
We utilize the structure information on exotic Neon isotopes calculated with the DRHBc theory described above as input for reaction calculations using a code based on the Glauber model~\cite{Abu-Ibrahim2003CPC151-369}. We specifically examine reaction~\textit{observables} that may indicate a halo structure in $^{31}$Ne, including a localized longitudinal momentum distribution in the breakup reaction, and a significant increase in both total reaction cross section and one-neutron removal cross section, in comparison to those quantities for neighboring Neon isotopes. Our reaction calculations here follow the same procedures described in our study at the spherical-shape limit~\cite{Zhang2021_arXiv2108.05609} and the references therein. 

\begin{figure}[htb]
\begin{center}
\includegraphics[width=0.5\textwidth]{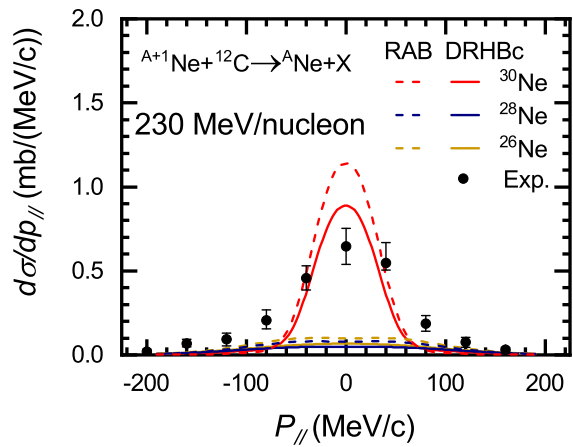}
\end{center}
\caption{
Calculated inclusive parallel momentum distribution of $^{26,28,30}$Ne residues after one-neutron removal from $^{27,29,31}$Ne bombarding a $^{12}$C target. Solid lines present results with DRHBc input, and dashed lines are for those with the spherical RAB model input~\cite{Zhang2021_arXiv2108.05609}. The experimental data for $^{30}$Ne are taken from Ref.~\cite{Nakamura2014_PRL112-142501}.
}
\label{fig4}
\end{figure}

First, we explore the longitudinal momentum distribution of $^{30}$Ne residues resulting from the bombardment of a $^{12}$C target with a beam of $^{31}$Ne. From the Heisenberg uncertainty principle, a peaked longitudinal momentum distribution of the residues of the inelastic breakup reaction implies a dilute spatial distribution in the $^{31}$Ne projectile before breakup. In Fig.~\ref{fig4}, we plot our Glauber model predictions (with DRHBc structure input) for the distributions for $^{26,28,30}$Ne residues after one-neutron removal breakup from $^{27,29,31}$Ne bombarding a $^{12}$C target at 230 MeV/nucleon. It is evident that momentum distributions for $^{26,28}$Ne are much flatter than that for $^{30}$Ne, consistent with the $^{31}$Ne projectile having a significantly larger size before breakup than $^{27,29}$Ne. Our predicted full width at half maximum (FWHM) distribution for the $^{30}$Ne residues is approximately 72 MeV/c, and compares well to the measured $^{30}$Ne distribution~\cite{Nakamura2014_PRL112-142501}. We note that our momentum distributions are almost the same for the NL1, NL3, and PK1 effective interactions.
For comparison, we also show in Fig.~\ref{fig4} the results from the spherical RAB model~\cite{Zhang2021_arXiv2108.05609}, which predicted a distribution that is more highly peaked than the experimental result. In summary, our predicted momentum distribution with deformed structure input is consistent with a halo structure for $^{31}$Ne and is in better agreement with experiment than results using a spherical structure model input for the Glauber model. 

\begin{figure}[htb]
\begin{center}
\includegraphics[width=0.5\textwidth]{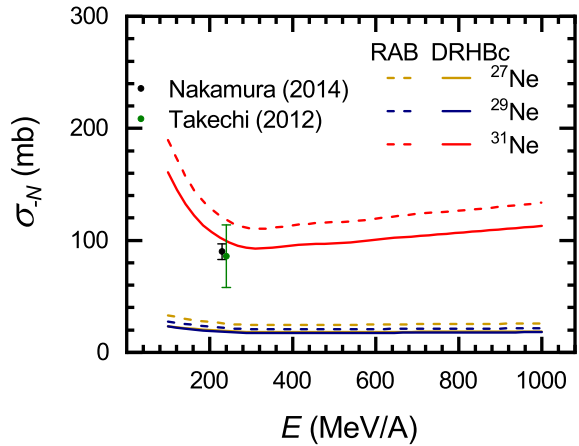}
\end{center}
\caption{
One-neutron removal cross section \textit{$\sigma_{-N}$} of $^{27,29,31}$Ne bombarding a $^{12}$C target as a function of incident energy per nucleon \textit{E}. Solid lines are results using DRHBc theory input, and dashed lines are those with spherical RAB model input~\cite{Zhang2021_arXiv2108.05609}. The experimental data are taken from Ref.~\cite{Nakamura2014_PRL112-142501, Takechi2012_PLB707-357}.
}
\label{fig5}
\end{figure}

Next, we examine the one-neutron removal cross sections $\sigma_{-N}$ for neutron-rich isotopes $^{27,29,31}$Ne. A significant increase in this cross section compared to neighboring nuclei indicates the presence of a neutron halo. We approximate the removal cross section by subtracting the relevant reaction cross sections, specifically as  \textit{$\sigma_R$}($^{A+1}$Ne) $-$ \textit{$\sigma_R$}($^{A}$Ne), where \textit{$\sigma_R$}($^{A}$Ne) is the reaction cross section to form the Neon isotope with mass $A$. Fig.~\ref{fig5} shows that our predicted one-neutron removal cross section prediction for $^{31}$Ne is a factor of 5--7 times larger than those of neighboring $^{29}$Ne and $^{27}$Ne.
This figure also shows that our result for $^{31}$Ne agrees with measurements~\cite{Nakamura2014_PRL112-142501, Takechi2012_PLB707-357}. We note that 
our predicted removal cross section for $^{31}$Ne decreases with energy near 240 MeV/nucleon; this agrees with the trend displayed by the \textit{central} values of the 2014 remeasurement at 230 MeV/nucleon~\cite{Nakamura2014_PRL112-142501} and the 2012 remeasurement at 240 MeV/nucleon~\cite{Takechi2012_PLB707-357}, although more measurements with lower uncertainties would be needed to show the decreasing trend. Finally, it is clear from Fig.~\ref{fig5} that the present results for $^{31}$Ne calculated with DRHBc theory input show a better agreement with the measured data than do the predictions using the spherical RAB structure model from Ref. ~\cite{Zhang2021_arXiv2108.05609}.
These results hold for the NL3 and PK1 effective interactions. In summary, our predicted one-neutron removal cross sections for $^{31}$Ne show a dramatic increase over neighboring Neon isotopes, clearly a signature of a neutron halo, and give better agreement than previous results using spherical model structure input. 

\begin{figure}[htb]
\begin{center}
\includegraphics[width=0.5\textwidth]{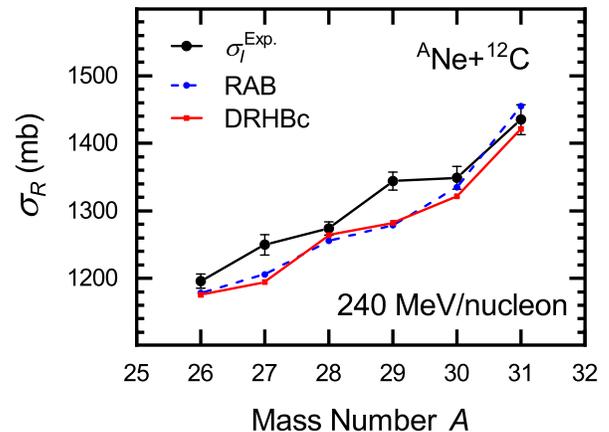}
\end{center}
\caption{
Reaction cross section \textit{$\sigma_R$} of $^{26-31}$Ne on a $^{12}$C target at 240 MeV/nucleon versus mass number \textit{A}.
Experimental data with uncertainties are taken from Ref.~\cite{Takechi2010_NPA834-412c},
the Glauber model predictions with DRHBc input (red solid line),
with spherical RAB structure input~\cite{Zhang2021_arXiv2108.05609} (blue dashed line).
%and with previous SHF input with SLy4 (green dotted line) and SkM$^*$ (pink dotted line) forces~\cite{horiuchi2012glauber}.
}
\label{fig6}
\end{figure}
 
Finally, we study trends in our Glauber model reaction cross section predictions of exotic Neon isotopes as a function of mass, and search for a signature of a halo in $^{31}$Ne. Fig.~\ref{fig6} shows our predictions of \textit{$\sigma_R$} for $^{26-31}$Ne bombarding a $^{12}$C target at 240 MeV/nucleon versus mass number \textit{A}. Also shown are the measured interaction cross sections from Ref.~\cite{Takechi2010_NPA834-412c}. Since the difference between reaction and interaction cross sections for inelastic reactions is negligible at 240 MeV/nucleon, we do not distinguish them in this study.

The measured cross sections increase by an average of $39 \pm 7$ mb/amu from $^{26}$Ne to $^{30}$Ne, but then exhibit a rapid rise of $86 \pm 39$ mb/amu from $^{30}$Ne to $^{31}$Ne, an increase in slope by a factor of $2.2 \pm 1.1$. Our predicted reaction cross sections using the  DRHBc input match this trend of a strong increase in reaction cross section for $^{31}$Ne: we show an increase of 37 mb/amu from $^{26}$Ne to $^{30}$Ne, which then increases by a factor of 2.7 from $^{30}$Ne to $^{31}$Ne. We note that this increase in cross section is consistent with our DRHBc prediction of an increase in the radius  
from $^{30}$Ne to $^{31}$Ne by a factor of 2.4 compared to that of neighboring nuclei. When using the NL3 and PK1 effective interactions, 
the total cross section predictions are approximately 40 mb ($3\%$) lower than predictions using the NL1 interaction.

For comparison, Fig.~\ref{fig6} also shows the predicted reaction cross section using the spherical RAB input~\cite{Zhang2021_arXiv2108.05609}. The trends are the same in both cases, with the spherical model slightly overpredicting the increase in cross section from $^{30}$Ne to $^{31}$Ne compared to Glauber model predictions using the DRHBc input. We also note that a previous study using SHF structure input with SLy4 and SkM$^*$ forces ~\cite{horiuchi2012glauber} well reproduce the reaction cross section trends from $^{26}$Ne to $^{30}$Ne, but do not reproduce the sudden increase of the slope from $^{30}$Ne to $^{31}$Ne that indicates a halo structure of $^{31}$Ne; for more details, see Ref.~\cite{Zhang2021_arXiv2108.05609}.

\section{SUMMARY}
\label{section 4}

We have described the exotic deformed nucleus $^{31}$Ne using an approach that combines structure and reaction theory. We utilized the fully-relativistic, microscopic deformed Hartree-Bogoliubov theory in continuum (DRHBc) to show that deformation and pairing correlations give rise to the halo structure in $^{31}$Ne with large amplitude (41.4$\%$) of $p$-wave configuration. Using the density distributions of the core nucleus and valence nucleon wave functions from this theory as input for the Glauber reaction model, we examine predictions of three possible halo signatures arising from the bombardment of a $^{12}$C target with a 240 (230) MeV/nucleon $^{31}$Ne beam. First, we predict an increase of reaction cross section from $^{30}$Ne to $^{31}$Ne that is 2.7 times larger than those of adjacent Neon isotopes, which is consistent with reaction measurements and with our calculated increase of the radius (2.4 times). 
Second, our calculated one-neutron removal cross sections are a factor of 5--7 higher than neighboring nuclei and have a decreasing trend with the energy near 240 MeV/nucleon, in  agreement with measurements within or nearly 1$\sigma$. Third, we calculate a narrow longitudinal momentum distribution of inelastic breakup products consistent with a dilute density distribution in coordinate space for the $^{31}$Ne projectile, whereas neighboring nuclei $^{27,29}$Ne show much flatter momentum distributions. Our predicted momentum distribution qualitatively agrees with the measured distribution and is improved over predictions using a spherical model to describe the $^{31}$Ne structure. 

The combination of the DRHBc structure theory, including pairing correlations and deformation effects, and the Glauber reaction model have now predicted all of the following halo characteristics for $^{31}$Ne:  large neutron- and matter-radii, reduced one-neutron separation energies, high $p$-orbital occupation probabilities, enhanced total reaction cross section, enhanced one-neutron removal cross section, and narrow breakup residue momentum distribution. 
Similar predictions of the momentum distribution and one-neutron removal cross sections are obtained when using the NL3 and PK1 effective interactions, but the total cross section predictions are about $3\%$ lower, compared to calculations using the NL1 interaction. For this reason, the NL1 effective interaction is to be preferred over the NL3 and PK1 interactions. Our approach with a deformed model, appropriate for $^{31}$Ne, gives an overall improvement over predictions that rely on a spherical structure model. We anticipate this self-consistent structure and reaction model approach can also be used to explain the properties of other halo nuclei.

\section{ACKNOWLEDGEMENTS}
\label{section 5}

This work was partially supported by the
National Natural Science Foundation of China under Grant
No.~11375022, No.~11775014, No.~11975237, and the Strategic Priority Research Program of the Chinese Academy of Sciences (Grant No. XDB34010000), and by the U.S. Department of Energy Office of Science, Office of Nuclear Physics, under Award Number DE-AC05-00OR22725.
The DRHBc calculations in this paper utilized the High-performance Computing Cluster of ITP-CAS and the SiGrid Supercomputing Center, Computer Network Information Center of CAS.

\bibliographystyle{model1a-num-names}
\bibliography{31Ne-DRHBc2021b}

\end{document}